\documentstyle[aps,prl,epsf,multicol,amsmath,graphics,psfrag,%
graphicx,pstcol,psfig]{revtex}

\begin{document}
\draft

\title{Renormalization approach to many-particle systems }

\author{K.~W.~Becker, A.~H\"{u}bsch, and T.~Sommer}
\address{Institut f\"{u}r Theoretische Physik,
  Technische Universit\"{a}t Dresden, D-01062 Dresden, Germany}

\date{\today}
\maketitle

\begin{abstract}
This paper presents a renormalization approach to many--particle
systems. By starting from a bare Hamiltonian 
${\cal H}= {\cal H}_0 +{\cal H}_1$ with an
unperturbed part ${\cal H}_0$ and a perturbation ${\cal H}_1$,
we define an effective Hamiltonian 
which has a band-diagonal shape with respect to the eigenbasis of 
${\cal H}_0$. This means that all 
transition matrix elements are suppressed which have energy differences 
larger than a given cutoff $\lambda$ that is smaller than the cutoff 
$\Lambda$ of the original Hamiltonian. 
This property resembles a recent flow 
equation approach on the basis of continuous unitary transformations. For 
demonstration of the method we discuss an exact solvable model, as well as 
the Anderson-lattice model where the well-known quasiparticle behavior of 
heavy fermions is derived.
\end{abstract}

\pacs{PACS numbers: 71.10.Fd, 71.27.+a, 75.20.Hr, 75.30.Mb}

\widetext
\begin{multicols}{2}
\narrowtext

\section{Introduction}

If one considers a many-particle problem one would like to diagonalize the 
Hamiltonian so that observables of interest can be calculated. However, 
only a few models can be solved explicitly so that often one has to use 
approximations or numerical approaches. Recently Wegner and coworkers 
\cite{Wegner1,Mielke} and G{\l}azek and Wilson \cite{Wilson1,Wilson2} 
proposed to solve the eigenvalue problem of a many-body system
by transforming the Hamiltonian in such a way 
that it becomes more and more diagonal. The method is based on the 
introduction of continuous unitary transformations 
and is formulated in 
terms of differential equations for the parameters of the 
Hamiltonian. These equations are called flow equations. 

In this paper we present a renormalization approach 
which is directly based on perturbation theory for the Hamiltonian 
instead of using a formulation in terms of differential equations.    
The starting point is a many--particle Hamiltonian 
${\cal H}={\cal H}_{0} + {\cal H}_{1}$. It is assumed that the 
eigenstates and eigenenergies of the unperturbed part ${\cal H}_{0}$ are 
known. The interaction ${\cal H}_{1}$ usually prevents the 
solution of the full eigenvalue problem of ${\cal H}$. In 
Sec.~\ref{pert}, we shall derive an effective 
Hamiltonian ${\cal H}_{\lambda}$ on the basis of a 
perturbational approach.  The obtained Hamiltonian 
${\cal H}_{\lambda}$ has no matrix elements (with 
respect to the unperturbed Hamiltonian ${\cal H}_{0}$) which belong 
to transitions larger than a given energy cutoff $\lambda$. In contrast to 
standard perturbation theory no vanishing energy denominators appear. In 
Sec.~\ref{Ren}, the result for ${\cal H}_{\lambda}$ will
be used to formulate a $\lambda$-dependent renormalization scheme  
by repeatedly applying a stepwise transformation on the 
effective Hamiltonian. In an 
infinitesimal formulation the renormalization of the Hamiltonian will be 
described by flow equations. At the end of this section a comparison 
of our approach  with those of Wegner \cite{Wegner1} and of G{\l}atzek and 
Wilson \cite{Wilson1,Wilson2} is given. For demonstration of the 
renormalization approach in Sec.~\ref{Fano} the exactly solvable 
Fano-Anderson model \cite{Anderson,Fano}  is discussed. This model 
describes two types of interacting electrons without correlations and
the exact result will be obtained in the framework 
of the renormalization approach. Finally, in
Sec.~\ref{Anderson} the application to the Anderson-lattice model 
\cite{Anderson} is discussed in order to derive the well-established  
quasiparticle behavior of heavy fermions. The conclusions in 
Sec.~\ref{Conc} conclude the paper.


\section{Effective Hamiltonian and perturbation theory}\label{pert}

Let us start from a decomposition of a given many-particle Hamiltonian 
${\cal H}$ into an unperturbed part ${\cal H}_{0}$ and 
into a perturbation ${\cal H}_{1}$
\begin{eqnarray}
  \label{1}
  {\cal H} &=& {\cal H}_{0} + \varepsilon {\cal H}_{1} =: H(\varepsilon) .
\end{eqnarray}
We assume that the eigenvalue problem of ${\cal H}_{0}$ is known
\begin{eqnarray}
  \label{2}
  {\cal H}_{0} |n\rangle &=& E^{(0)}_n |n\rangle .
\end{eqnarray}
${\cal H}_{1}$ is the interaction. Its presence usually prevents a solution of 
the eigenvalue problem of the full Hamiltonian.
The parameter $\varepsilon$ accounts for the order of perturbation processes
discussed below. Let us define a projection operator ${\bf P}_{\lambda}$ by
\begin{eqnarray}
  \label{3}
  {\bf P}_{\lambda}A &=& \sum_{|E_n^{(0)} - E_m^{(0)}| \leq \lambda}
  |n\rangle \langle m| \, \langle n| A  |m\rangle .
\end{eqnarray}
Here, ${\bf P}_{\lambda}$ is a superoperator acting on usual 
operators $A$ of the unitary space. It projects on that part of $A$ which is
formed by all dyads $|n\rangle \langle m|$ with energy differences
$|E_n^{(0)} - E_m^{(0)}|$ less or equal to a given cutoff $\lambda$,
where $\lambda$ is  smaller 
than the cutoff $\Lambda$ of the original model. Note that in Eq. \eqref{3} 
neither  $|n\rangle$ nor $|m\rangle$ have to be low-energy eigenstates of 
$H_0$. However, their energy difference has to be restricted to 
values $\leq \lambda$. Furthermore, it is useful to define the projection 
operator 
\begin{eqnarray}
  \label{4}
  {\bf Q}_{\lambda} &=& {\bf 1} - {\bf P}_{\lambda}
\end{eqnarray}
which is orthogonal to ${\bf P}_{\lambda}$. ${\bf Q}_{\lambda}$ projects on 
high-energy transitions larger than the cutoff $\lambda$.

The goal of the present method is to transform the initial Hamiltonian 
${\cal H}$
(with a large energy cutoff $\Lambda$) into an effective Hamiltonian 
${\cal H}_{\lambda}$ which has no matrix elements belonging to transitions 
larger than $\lambda$. This is achieved by a unitary 
transformation so that the effective Hamiltonian will have the same 
eigenspectrum as the original Hamiltonian ${\cal H}$. However, 
as it will turn out, it is especially suitable  
to describe the low-energy excitations of the system.  
${\cal H}_{\lambda}$ is defined by  
\begin{eqnarray}
  \label{5}
  {\cal H}_{\lambda} &=&
  e^{X_{\lambda}} \, {\cal H} \, e^{-X_{\lambda}} .
\end{eqnarray}
The generator $X_{\lambda}$ of the transformation has to be 
anti--Hermitian, $X^{\dagger}_{\lambda}=-X_{\lambda}$, so that  
${\cal H}_{\lambda}$ is Hermitian for any $\lambda$. We look for an 
appropriate  generator $X_{\lambda}$ so that ${\cal H}_{\lambda}$
has no matrix elements belonging to transitions larger than $\lambda$. This 
means that the following condition  
\begin{eqnarray}
  \label{6}
  {\bf Q}_{\lambda}{\cal H}_{\lambda} &=& 0
\end{eqnarray}
has to be fulfilled. Equation \eqref{6}  will be used below to 
specify $X_{\lambda}$. 

The expression \eqref{5} for the effective Hamiltonian 
${\cal H}_{\lambda}$ can we expanded with respect to $X_{\lambda}$
\begin{eqnarray}
  {\cal H}_{\lambda} &=&
  {\cal H} + 
  \left[
    X_{\lambda}, {\cal H}
  \right] + 
  \frac{1}{2!}
  \left[
    X_{\lambda},
    \left[
      X_{\lambda}, {\cal H}
    \right]
  \right]\nonumber\\
  &&
  \label{7}
  + \frac{1}{3!}
  \left[
    X_{\lambda},
    \left[
      X_{\lambda},
      \left[
        X_{\lambda}, {\cal H}
      \right]
    \right]
  \right] + \dots \; .
\end{eqnarray}
By assuming that $X_{\lambda}$ can be written as a power series in the 
perturbation parameter $\varepsilon$ 
\begin{eqnarray}
  \label{8}
  X_{\lambda} &=& 
  \varepsilon X_{\lambda}^{(1)} + \varepsilon^{2} X_{\lambda}^{(2)} + 
  \varepsilon^{3} X_{\lambda}^{(3)} + \dots
\end{eqnarray}
the effective Hamiltonian ${\cal H}_{\lambda}$ can be expanded in a 
power series in $\varepsilon$ as well 
\begin{eqnarray}
  \label{9}
  {\cal H}_{\lambda} &=& 
  {\cal H}_{0} + 
  \varepsilon
  \left\{
    {\cal H}_{1} + 
    \left[
      X_{\lambda}^{(1)}, {\cal H}_{0}
    \right]
  \right\} \\
  &&
  + \varepsilon^{2}
  \left\{
    \left[
      X_{\lambda}^{(1)}, {\cal H}_{1}
    \right] + 
    \left[
      X_{\lambda}^{(2)}, {\cal H}_{0}
    \right]
  \right. \nonumber \\
  &&
  \qquad\qquad\left.
    + \frac{1}{2!}
    \left[
      X_{\lambda}^{(1)},
      \left[
        X_{\lambda}^{(1)}, {\cal H}_{0}
      \right]
    \right]
  \right\} + {\cal O}(\varepsilon^{3}) \nonumber.
\end{eqnarray}
The different contributions 
$X_{\lambda}^{(n)}$ to the generator of the unitary transformation \eqref{5} 
can be successively determined from Eq.~\eqref{6}. One finds 
\begin{eqnarray}
  \label{10}
  {\bf Q}_{\lambda} X_{\lambda}^{(1)} &=& 
  \frac{1}{{\bf L}}_{0}
  \left(
    {\bf Q}_{\lambda} {\cal H}_{1}
  \right) , \\[3ex]
  \label{11}
  {\bf Q}_{\lambda} X_{\lambda}^{(2)} &=&
  - \frac{1}{2 {\bf L}_{0} } {\bf Q}_{\lambda}
  \left[
    ( {\bf Q}_{\lambda}{\cal H}_{1}), \frac{1}{{\bf L}_{0}}
    ( {\bf Q}_{\lambda}{\cal H}_{1})
  \right]\\
  &&
  - \frac{1}{{\bf L}_{0}}
  {\bf Q}_{\lambda}
  \left[
    ({\bf P}_{\lambda}{\cal H}_{1}), \frac{1}{ {\bf L}_{0} } 
    ({\bf Q}_{\lambda}{\cal H}_{1})
  \right]\nonumber , 
\end{eqnarray}
and similar expression in higher order in $\varepsilon$.  
Here ${\bf L}_{0}$ denotes the unperturbed Liouville operator which is 
defined by ${\bf L}_{0}A=[{\cal H}_{0},A]$ for any operator variable $A$. 
Note that by Eqs. \eqref{10}, \eqref{11}, and the equivalent equations for the 
higher orders in $\varepsilon$ the generators $X_{\lambda}^{(n)}$  are not 
completely fixed. The reason is that the parts
${\bf P}_{\lambda} X_{\lambda}^{(n)}$ of  $X_{\lambda}^{(n)}$ are not 
fixed by Eq. \eqref{6}  and can still 
be chosen arbitrarily. In the following we use for convenience 
\begin{eqnarray}
  \label{12}
  {\bf P}_{\lambda} X_{\lambda}^{(1)} &=& 
  {\bf P}_{\lambda} X_{\lambda}^{(2)} \,=\, 0 .
\end{eqnarray} 
Inserting Eqs. \eqref{10} and \eqref{11} into Eq. \eqref{9} one finds for the 
effective Hamiltonian ${\cal H}_{\lambda}$ up to second order in ${\cal H}_1$ 
\begin{eqnarray}
  \label{13}
  {\cal H}_{\lambda} &=&
  {\cal H}_{0} + {\bf P}_{\lambda}{\cal H}_{1} - 
  \frac{1}{2} {\bf P}_{\lambda}
  \left[
    ( {\bf Q}_{\lambda}{\cal H}_{1}), \frac{1}{{\bf L}_{0}}
    ( {\bf Q}_{\lambda}{\cal H}_{1})
  \right] \\
  && 
  - {\bf P}_{\lambda}
  \left[
    ({\bf P}_{\lambda}{\cal H}_{1}), \frac{1}{ {\bf L}_{0} } 
    ({\bf Q}_{\lambda}{\cal H}_{1})
  \right],\nonumber
\end{eqnarray}
where $\varepsilon$ was set equal to $1$. 
Note that the perturbation expansion \eqref{13} can  
easily be extended to higher orders in $\varepsilon$. One 
should also note that Eq. \eqref{13} automatically guarantees the correct 
size dependence of the Hamiltonian due to the commutators appearing 
in Eq. \eqref{13}. 

In the following section we will use Eq. \eqref{13}
to establish a renormalization approach to  many-particle systems 
by successively reducing the cutoff energy $\lambda$ to smaller 
and smaller values.  
In the limit $\lambda \rightarrow 0$ 
expression \eqref{13} provides an expression 
for an effective Hamiltonian that acts on a possibly 
degenerate ground-state of the unperturbed Hamiltonian.  
For this case one can 
also show that Eq. \eqref{13} reduces to an expression 
derived from usual perturbation theory \cite{Takahashi}. 
A systematic extension of Eq.~\eqref{13} to higher-order
perturbation theory 
can also be used to provide a simple scheme for an algebraic  
evaluation of higher terms of ${\cal H}_{(\lambda=0)}$ 
by use of a computer \cite{Sykora}. Such a  scheme does not require special 
properties of the eigenvalue spectrum of the unperturbed Hamiltonian 
${\cal H}_{0}$. In particular, an equidistant spectrum is not needed. 
This property was implied for a perturbation expansion \cite{Knetter} 
based on Wegner's flow equation method.

\section{Renormalization approach}\label{Ren}

In this section we will discuss the elimination procedure for the
interaction ${\cal H}_1$. However, instead of transforming the Hamiltonian
in one step as was done in the preceding section the transformation
will be performed successively.
Or more formally spoken, instead of applying the elimination of
high-energy excitations in one step a sequence of
stepwise transformations is used in order to obtain an effectively 
diagonal model. In an infinitesimal formulation the renormalization of the 
coupling constants will be described by flow equations. In order to find these
equations some approximations will be necessary. The goal of the method is to
transform the initial Hamiltonian (with a large cutoff $\Lambda$) into an
effective Hamiltonian which  possesses only coupling terms between states with
an energy difference less or equal to $\lambda$. Thereby, the method yields
renormalization equations as function of the cutoff $\lambda$.

Let us start from the renormalized Hamiltonian
\begin{eqnarray}
  \label{14}
  {\cal H}_{\lambda} &=& {\cal H}_{0,\lambda} + {\cal H}_{1,\lambda}
\end{eqnarray}
after all excitations with energy differences larger than $\lambda$ have been
eliminated. In the next step an additional renormalization of 
${\cal H}_{\lambda}$ to a new 
Hamiltonian ${\cal H}_{(\lambda-\Delta\lambda)}$  is done
by eliminating  all excitations within the energy shell between 
the cutoff 
$\lambda$ and a somewhat smaller energy cutoff  
$(\lambda-\Delta\lambda)$. The new Hamiltonian 
${\cal H}_{(\lambda-\Delta\lambda)}$ will  be calculated by use of 
the perturbation theory discussed above. One finds 
\begin{eqnarray}
  \label{15}
  \lefteqn{
    {\cal H}_{(\lambda-\Delta\lambda)} \,=\,
  }&& \\
  &=&
  {\bf P}_{(\lambda-\Delta\lambda)} {\cal H}_{\lambda} \nonumber\\
  && 
  - \frac{1}{2} {\bf P}_{(\lambda-\Delta\lambda)}
  \left[
    ( {\bf Q}_{(\lambda-\Delta\lambda)}{\cal H}_{1,\lambda}), 
    \frac{1}{{\bf L}_{0,\lambda}}
    ( {\bf Q}_{(\lambda-\Delta\lambda)}{\cal H}_{1,\lambda})
  \right] \nonumber\\
  &&  
  - {\bf P}_{(\lambda-\Delta\lambda)}
  \left[
    ({\bf P}_{(\lambda-\Delta\lambda)}{\cal H}_{1,\lambda}), 
    \frac{1}{ {\bf L}_{0,\lambda} } 
    ({\bf Q}_{(\lambda-\Delta\lambda)}{\cal H}_{1,\lambda})
  \right] . \nonumber
\end{eqnarray} 
${\bf L}_{0,\lambda}$ is now the Liouville operator 
with respect to the $\lambda$-dependent 
unperturbed Hamiltonian ${\cal H}_{0,\lambda}$. 
As is obvious from Eq.~\eqref{15} the 
second-order term on the right-hand side (rhs) may be divided into two parts. 
The first one connects eigenstates of ${\cal H}_{0,\lambda}$ with the same 
energy. This part commutes with ${\cal H}_{0,\lambda}$ and can therefore be 
considered as renormalization of the unperturbed Hamiltonian
\begin{eqnarray}
  \label{16}
  \lefteqn{
    {\cal H}_{0,(\lambda-\Delta\lambda)} - {\cal H}_{0,\lambda} \,=\,
  } && \\
  &&
  - \frac{1}{2} \, {\bf P}_{0}
  \left[
    ( {\bf Q}_{(\lambda-\Delta\lambda)}{\cal H}_{1,\lambda}), 
    \frac{1}{{\bf L}_{0,\lambda}}
    ( {\bf Q}_{(\lambda-\Delta\lambda)}{\cal H}_{1,\lambda})
  \right] . \nonumber
\end{eqnarray}
In contrast the second part connects eigenstates of ${\cal H}_{0,\lambda}$ 
with different energies. It therefore represents a renormalization 
of the interaction part of the Hamiltonian 
\begin{eqnarray}
  \label{17}
  \lefteqn{
    {\cal H}_{1,(\lambda-\Delta\lambda)} - 
    {\bf P}_{(\lambda-\Delta\lambda)}{\cal H}_{1,\lambda}
    \,=\,
  } && \\
  &=&
  - {\bf P}_{(\lambda-\Delta\lambda)}
  \left[
    ({\bf P}_{(\lambda-\Delta\lambda)}{\cal H}_{1,\lambda}), 
    \frac{1}{ {\bf L}_{0,\lambda} } 
    ({\bf Q}_{(\lambda-\Delta\lambda)}{\cal H}_{1,\lambda})
  \right]  \nonumber \\ 
  && 
  -\frac{1}{2} 
  \left(
    {\bf P}_{(\lambda-\Delta\lambda)} - {\bf P}_{0}
  \right) \nonumber\\
  &&
  \qquad
  \times\left[
    ( {\bf Q}_{(\lambda-\Delta\lambda)}{\cal H}_{1,\lambda}), 
    \frac{1}{{\bf L}_{0,\lambda}}
    ( {\bf Q}_{(\lambda-\Delta\lambda)}{\cal H}_{1,\lambda})
  \right]  \nonumber  \\
&& \nonumber \\
&\approx& 
  - {\bf P}_{(\lambda-\Delta\lambda)}
  \left[
    ({\bf P}_{(\lambda-\Delta\lambda)}{\cal H}_{1,\lambda}), 
    \frac{1}{ {\bf L}_{0,\lambda} } 
    ({\bf Q}_{(\lambda-\Delta\lambda)}{\cal H}_{1,\lambda})
  \right] .\nonumber  \\
&& \nonumber  
\end{eqnarray}
We assume that $\Delta\lambda$ is small. Therefore,  only the 
mixed term, i.e., the second part on the rhs 
of Eq.~\eqref{17},  contributes to the renormalization of 
${\cal H}_{1,\lambda}$. Thus,  
Eqs. \eqref{16}  and \eqref{17} describe the 
renormalization of the unperturbed and of the interaction
part of the Hamiltonian. They represent the main result of our  
theoretical formalism. In the limit $\Delta\lambda \rightarrow 0$, i.e.~for 
vanishing shell width, these equation can be used to derive 
differential equations for the dependence of the 
Hamiltonian on the cutoff energy $\lambda$. The resulting 
equations for the parameters of the Hamiltonian are called flow equations.
Their solution depend on the initial values of the parameters of
the Hamiltonian and determine to final Hamiltonian in the 
limit $\lambda \rightarrow 0 $. Note that for $\lambda \rightarrow 0$ 
the resulting Hamiltonian only consists of the unperturbed part 
${\cal H}_{0,(\lambda \rightarrow 0)}$, so that an effectively diagonal 
Hamiltonian is obtained. The interaction 
${\cal H}_{1,(\lambda\rightarrow 0)}$
completely vanishes since it is completely used up in the renormalization
procedure. The detailed structure of the flow equations strongly
depends on the system under consideration. However, one can easily 
write down these equations in matrix representation with 
respect to the unperturbed Hamiltonian (compare Appendix A).     

Next we discuss the relation between the present renormalization 
approach and Wegner's flow equation method \cite{Wegner1}. The approach 
presented here is also based on the idea to formulate flow equations for the 
Hamiltonian. However, there are substantial differences between both methods: 
i) The present approach starts directly from the renormalization of the 
Hamiltonian and not from continuous unitary transformations in differential 
form.
ii) Moreover, the band diagonal form of the transformed Hamiltonian results 
from a physically motivated projection operator on low-energy transitions 
instead of introducing cutoff functions to restrict the Hamiltonian matrix. 
iii) The result of the renormalization approach \eqref{16} and \eqref{17} is 
directly formulated in terms of an operator relation. Wegner's flow equation 
approach \cite{Wegner1} is usually written in matrix representation though it 
is also possible to give a formulation in terms of operators \cite{Grote}.
iv) As was discussed in Sec.~\ref{pert}, there is a direct connection between 
the renormalization method and usual perturbation theory.

\section{The exact solvable Fano-Anderson model}\label{Fano}

In this section we illustrate the renormalization approach for the case of a 
simple model, i.e., we apply the method to the exact solvable Fano-Anderson 
model \cite{Anderson,Fano}. This model is given by
\begin{eqnarray}
  \label{18}
  {\cal H} &=& {\cal H}_{0} + {\cal H}_{1} ,
\end{eqnarray}
\begin{eqnarray}
  {\cal H}_{0} &=&
  \varepsilon_{f} \sum_{i,m} f^{\dagger}_{im} f_{im} +
  \sum_{{\bf k},m} 
  \varepsilon_{{\bf k}} \, c^{\dagger}_{{\bf k}m} c_{{\bf k}m} \nonumber\\
  &=&
  \sum_{{\bf k},m}
  \left(
    \varepsilon_{f} \, f^{\dagger}_{{\bf k}m} f_{{\bf k}m} +
    \varepsilon_{{\bf k}} \, c^{\dagger}_{{\bf k}m} c_{{\bf k}m}
  \right) , \nonumber
\end{eqnarray}
\begin{eqnarray}
  {\cal H}_{1} &=&
  \frac{1}{\sqrt{N}} \sum_{{\bf k},i,m} V_{{\bf k}}
  \left(
    f_{im}^{\dagger}c_{{\bf k}m} \, e^{{\rm i}{\bf k}{\bf R}_{i}} + {\rm h.c.}
  \right) \nonumber \\
  &=& 
  \sum_{{\bf k},m} V_{{\bf k}}
  \left(
    f_{{\bf k}m}^{\dagger}c_{{\bf k}m} + c_{{\bf k}m}^{\dagger}f_{{\bf k}m}
  \right) . \nonumber
\end{eqnarray}
A possible realization of the model is a periodic system of localized
$f$ electrons which interact with conducting electrons thereby
neglecting correlation effects. The index $i$ denotes the $f$ sites,
${\bf k}$ is the wave vector, and $V_{{\bf k}}$ describes the hybridization
between conduction and localized electrons. The excitation energies
$\varepsilon_{{\bf k}}$ and $\varepsilon_{f}$ for conduction and localized 
electrons are measured from the chemical potential $\mu$. As a further 
simplification, both types of electrons are assumed to have the same angular 
momentum index $m$ with values $m=1 \dots \nu_{f}$. Of course, the model is 
easily solved and leads to two hybridized bands 
\begin{eqnarray}
  \label{19}
  {\cal H} &=&  
  \sum_{{\bf k},m} \omega_{{\bf k}}^{(\alpha)} 
  \alpha_{{\bf k}m}^{\dagger} \alpha_{{\bf k}m} + 
  \sum_{{\bf k},m} \omega_{{\bf k}}^{(\beta)} 
  \beta_{{\bf k}m}^{\dagger} \beta_{{\bf k}m} , \\[2ex]
  \omega_{{\bf k}}^{(\alpha,\beta)} &=& 
  \frac{\varepsilon_{{\bf k}} + \varepsilon_f}{2}
  \pm \frac{1}{2} \sqrt{(\varepsilon_{{\bf k}} - 
  \varepsilon_f)^2 + 4 |V_{{\bf k}}|^2}
  \nonumber
\end{eqnarray}
with  eigenmodes $\alpha_{{\bf k}m}^{\dagger}$ and 
$\beta_{{\bf k}m}^{\dagger}$ given by certain linear combinations of 
$c_{{\bf k}m}^{\dagger}$ and $f_{{\bf k}m}^{\dagger}$, 
i.e., 
\begin{eqnarray}
  \alpha_{{\bf k}m}^{\dagger} &=& 
  u_{{\bf k}} \, f_{{\bf k}m}^{\dagger} + v_{{\bf k}} \, 
  c_{{\bf k}m}^{\dagger}, \quad 
  \beta_{{\bf k}m}^{\dagger} \,=\, 
  - v_{{\bf k}} \, f_{{\bf k}m}^{\dagger} + u_{{\bf k}} \, 
  c_{{\bf k}m}^{\dagger} , \nonumber
\end{eqnarray}
\begin{eqnarray}
  |u_{{\bf k}}|^{2} &=& 
  \frac{1}{2} - 
  \frac{1}{2}
  \frac{\varepsilon_{{\bf k}}-\varepsilon_{f}}
  {
    \sqrt{
      \left(
        \varepsilon_{{\bf k}}-\varepsilon_{f}
      \right)^{2} + 4 |V_{{\bf k}}|^{2}
    }
  } ,\nonumber \\
  \label{20}
  |v_{{\bf k}}|^{2} &=&
  \frac{1}{2} + 
  \frac{1}{2}
  \frac{\varepsilon_{{\bf k}}-\varepsilon_{f}}
  {
    \sqrt{
      \left(
        \varepsilon_{{\bf k}}-\varepsilon_{f}
      \right)^{2} + 4 |V_{{\bf k}}|^{2}
    }
  } . 
\end{eqnarray} 

In the renormalization approach we are integrating out particle-hole 
excitations of conduction and $f$ electrons described by the hybridization 
${\cal H}_1$. We expect to obtain finally an effectively free model. 

The starting point of the method is a  renormalized Hamiltonian
${\cal H}_{\lambda}$ which is obtained after all excitations with energies 
larger than a given cutoff  $\lambda$ have been eliminated. Due to the result 
of the preceeding section it should have the following form 
\begin{eqnarray}
  \label{21}
  {\cal H}_{\lambda} &=& {\cal H}_{0,\lambda} + {\cal H}_{1,\lambda} , \\[2ex]
  {\cal H}_{0, \lambda} &=&
   \sum_{{\bf k},m}
  \left(
    \varepsilon_{{\bf k},\lambda}^{f} \, f^{\dagger}_{{\bf k}m} f_{{\bf k}m} +
    \varepsilon_{{\bf k},\lambda}^{c} \, c^{\dagger}_{{\bf k}m} c_{{\bf k}m}   
  \right) , 
  \nonumber \\
 {\cal H}_{1,\lambda}
  &=&  
  \sum_{
    \genfrac{}{}{0pt}{1}{
      \genfrac{}{}{0pt}{1}{{\bf k},m}{
        \left|
          \varepsilon_{{\bf k},\lambda}^{c} - \varepsilon_{{\bf k},\lambda}^{f}
        \right|
        \leq \lambda
      }
    }{}
  }
  V_{{\bf k}}
  \left(
    f_{{\bf k}m}^{\dagger}c_{{\bf k}m} + c_{{\bf k}m}^{\dagger}f_{{\bf k}m}
  \right) =  {\bf P}_{\lambda} \, {\cal H}_{1} .
 \nonumber
\end{eqnarray}

As it turns out no renormalization of the hybridization $ V_{\bf k}$ occurs so 
that $V_{\bf k}$ is assumed to be $\lambda$ independent from the beginning. 
However, the one-particle energies $\varepsilon_{{\bf k},\lambda}^{c}$ and 
$\varepsilon_{{\bf k},\lambda}^{f}$ will depend on the energy cutoff 
$\lambda$ and moreover $\varepsilon_{{\bf k},\lambda}^{f}$ depend on the wave 
vector ${\bf k}$. Next,  we evaluate the new Hamiltonian 
${\cal H}_{0,(\lambda-\Delta\lambda)}$ by projecting out all excitations 
between $\lambda -\Delta\lambda$ and $\lambda$ [compare \eqref{16}]. By 
inserting Eq. \eqref{21} into Eq. \eqref{16}  we obtain
\begin{eqnarray}
  \label{22}
  \lefteqn{
    \delta{\cal H}_{0}(\lambda,\Delta\lambda) \,=\, 
    {\cal H}_{0,(\lambda-\Delta\lambda)} - {\cal H}_{0,\lambda}
  }&& \\
  &=& 
  {\sum_{{\bf k},m}}^{\prime}
  \left\{
    - \frac{ |V_{{\bf k}}|^{2} }
    { \varepsilon_{{\bf k},\lambda}^{c} - \varepsilon_{{\bf k},\lambda}^{f} }
    + \dots
  \right\}
  \left[
    f_{{\bf k}m}^{\dagger}f_{{\bf k}m} - c_{{\bf k}m}^{\dagger}c_{{\bf k}m}
  \right] \nonumber
\end{eqnarray}
where the Fermion commutation relations were used. The prime $\prime$ above 
the sum indicates that the condition
$
  (\lambda-\Delta\lambda) <
  |
    \varepsilon_{{\bf k},\lambda}^{c} - \varepsilon_{{\bf k},\lambda}^{f}
  |
  \leq \lambda
$
has to be fulfilled. Note that in Eq.~\eqref{22} only second-order 
contributions are explicitly given. Higher-order terms are indicated by dots 
$\dots$ . The renormalization equations for the one-particle energies are 
easily found from Eq. \eqref{22} and read 
\begin{eqnarray}
  \varepsilon_{{\bf k},(\lambda-\Delta\lambda)}^{f} - 
  \varepsilon_{{\bf k},\lambda}^{f} 
  &=&
  \left\{
    - \frac{ |V_{{\bf k}}|^{2} }
    { \varepsilon_{{\bf k},\lambda}^{c} - \varepsilon_{{\bf k},\lambda}^{f} }
    + \dots
  \right\} \;
  \Theta(\lambda,\Delta\lambda),\nonumber\\[-2ex]
  \label{23}
  && \\
  \varepsilon_{{\bf k},(\lambda-\Delta\lambda)}^{c} - 
  \varepsilon_{{\bf k},\lambda}^{c}
  &=&
  - \left\{
    - \frac{ |V_{{\bf k}}|^{2} }
    { \varepsilon_{{\bf k},\lambda}^{c} - \varepsilon_{{\bf k},\lambda}^{f} }
    + \dots
  \right\} \;
  \Theta(\lambda,\Delta\lambda)\nonumber\\[-2ex]
  \label{24}
  &&
\end{eqnarray}
where
\begin{eqnarray}
  \label{24a}
  \Theta(\lambda,\Delta\lambda) &=& 
  \Theta\left(
    |
      \varepsilon_{{\bf k},\lambda}^{c} - \varepsilon_{{\bf k},\lambda}^{f}
    | - (\lambda-\Delta\lambda)
  \right) \nonumber\\
  &&
  -  \Theta\left(
    |
      \varepsilon_{{\bf k},\lambda}^{c} - \varepsilon_{{\bf k},\lambda}^{f}
    | - \lambda
  \right).
\end{eqnarray}
To solve these equations we subtract Eq. \eqref{23} from Eq. \eqref{24}, 
multiply both sides by 
$(\varepsilon_{{\bf k},\lambda}^{c} - \varepsilon_{{\bf k},\lambda}^{f})$, and 
divide by $\Delta\lambda$. By performing the limit 
$\Delta\lambda\rightarrow 0$ we obtain
\begin{eqnarray}
  \label{25}
  \frac{d}{d\lambda}
  \left(
    \varepsilon_{{\bf k},\lambda}^{c} - \varepsilon_{{\bf k},\lambda}^{f}
  \right)^{2}
  &=&
  - 4
  \left|
    V_{{\bf k}}
  \right|^{2} \;
  \lim_{\Delta\lambda\rightarrow 0}
  \frac{\Theta(\lambda,\Delta\lambda)}{\Delta\lambda}.
\end{eqnarray}
Note that Eq. \eqref{25} includes all renormalization contributions of 
higher orders in the interaction. It turns out that the solution of 
Eq. \eqref{25} shows a step like behavior and reads
\begin{eqnarray}
  \label{25a}
  | \varepsilon_{{\bf k},\lambda}^{c}-\varepsilon_{{\bf k},\lambda}^{f} | &=& 
  | \varepsilon_{{\bf k}} - \varepsilon_{f} | + 
  C_{{\bf k}} 
  \Theta ( | \varepsilon_{{\bf k}} - \varepsilon_{f} | - \lambda ) \\
  \mbox{with} \qquad 
  C_{{\bf k}} &=& \sqrt{(\varepsilon_{{\bf k}} - \varepsilon_{f})^{2} + 
  4|V_{{\bf k}}|^{2}}
  - | \varepsilon_{{\bf k}} - \varepsilon_{f} | . \nonumber
\end{eqnarray}
To prove this result one has to insert Eq. \eqref{25a} into Eq. \eqref{25} 
which gives 
$
  -4|V_{\bf k}|^{2} 
  \delta(|\varepsilon_{\bf k}-\varepsilon_{f}|-\lambda)
$
for both sides. Note that the evaluation of the rhs. of Eq. \eqref{25} 
together with Eq. \eqref{24a} does not give 
$
  -4|V_{\bf k}|^{2} 
  \delta(
    |\varepsilon_{{\bf k},\lambda}^{c}-\varepsilon_{{\bf k},\lambda}^{f}|
    -\lambda
  )
$
as would be expected of first glance. This follows from the step like 
renormalization of 
$|\varepsilon_{{\bf k},\lambda}^{c}-\varepsilon_{{\bf k},\lambda}^{f}|$, 
described by Eq. \eqref{25a}.  

For $\lambda\rightarrow 0$ one finds
\begin{eqnarray}
  \label{26}
  \left(
    {\tilde\varepsilon}_{{\bf k}}^{c} - {\tilde\varepsilon}_{{\bf k}}^{f}
  \right)^{2}
  &=&
  \left(
    \varepsilon_{{\bf k}} - \varepsilon_{f}
  \right)^{2}
  + 4
  \left|
    V_{{\bf k}}
  \right|^{2} 
\end{eqnarray}
where $\tilde{\varepsilon}_{{\bf k}}^{c}$ and 
$\tilde{\varepsilon}_{{\bf k}}^{f}$ denote the one-particle energies at 
$\lambda \rightarrow 0$,
\begin{eqnarray}
  \label{27}
  {\tilde\varepsilon}_{{\bf k}}^{c} &=& 
  \varepsilon_{{\bf k},(\lambda\rightarrow 0)}^{c}, 
  \hspace*{1cm} 
  {\tilde\varepsilon}_{{\bf k}}^{f}= 
  \varepsilon_{{\bf k},(\lambda\rightarrow 0)}^f. 
\end{eqnarray}
On the other hand, it is easily seen that the summation of 
Eqs. \eqref{23} and \eqref{24} leads to 
\begin{eqnarray}
  \label{29}
  {\tilde\varepsilon}_{{\bf k}}^{c} + {\tilde\varepsilon}_{{\bf k}}^{f} &=&
  \varepsilon_{{\bf k}} + \varepsilon_{f} .
\end{eqnarray}
The system of Eqs. \eqref{26} and \eqref{29} can easily be 
solved and gives 
\begin{eqnarray}
  \label{30}
  {\tilde\varepsilon}_{{\bf k}}^{c} &=&
  \frac{\varepsilon_{{\bf k}} + \varepsilon_{f}}{2} +
  \frac{1}{2}
  \sqrt{
    \left(
      \varepsilon_{{\bf k}} - \varepsilon_{f}
    \right)^{2}
    + 4 \left| V_{{\bf k}} \right|^{2} ,
  } \\
&& \nonumber \\
  \label{31}
  {\tilde\varepsilon}_{{\bf k}}^{f} &=&
  \frac{\varepsilon_{{\bf k}} + \varepsilon_{f}}{2} -
  \frac{1}{2}
  \sqrt{
    \left(
      \varepsilon_{{\bf k}} - \varepsilon_{f}
    \right)^{2}
    + 4 \left| V_{{\bf k}} \right|^{2}
  } .
\end{eqnarray}
Note that Eq. \eqref{31} agrees with the exact solution Eq. \eqref{19}. As a 
result we have obtained the following free model
\begin{eqnarray}
  \label{31a} 
  {\cal H}_{(\lambda\rightarrow 0)} &=& 
  \sum_{{\bf k},m} 
  \left(
    \tilde{\varepsilon}_{\bf k}^{c} \,c_{{\bf k}m}^{\dagger} c_{{\bf k}m}
    + \tilde{\varepsilon}_{\bf k}^{f} \, f_{{\bf k}m}^{\dagger} f_{{\bf k}m} 
  \right) .
\end{eqnarray}

Note that the interaction ${\cal H}_{1,\lambda}$ vanishes for 
$\lambda \rightarrow 0$ since it is completely used to renormalize the 
one-particle energies $\tilde{\varepsilon}_{\bf k}^{c}$ and 
$\tilde\varepsilon_{\bf k}^{f}$. The result Eqs. \eqref{31a} and \eqref{31} 
can be used 
to evaluate expectation values. Since ${\cal H}_{(\lambda \rightarrow 0)}$ 
emerged from the original model ${\cal H}$ by an unitary transformation, the 
free energie can also be calculated from ${\cal H}_{(\lambda \rightarrow 0)}$
\begin{eqnarray}
  \label{32}
  F &=& 
  - \frac{1}{\beta} \ln {\rm Tr} \, 
  e^{-\beta {\cal H}}
  \;=\;
  - \frac{1}{\beta} \ln {\rm Tr} \, 
  e^{-\beta {\cal H}_{(\lambda\rightarrow 0)}} =: F_{(\lambda \rightarrow 0)} .
  \nonumber\\
  &&
\end{eqnarray}  
Then, the $f$ electron occupation number is found from 
$F_{(\lambda \rightarrow 0)}$ by functional derivative
\begin{eqnarray}
  \label{33}
  \langle f_{{\bf k}m}^{\dagger}f_{{\bf k}m} \rangle &=& 
  \frac{1}{N}\frac{\partial F}{\partial \varepsilon_{f}} \;=\;
  \frac{1}{N}\frac{\partial F_{(\lambda \rightarrow 0)}}
  {\partial \varepsilon_{f}} \\
 &=& \frac{1}{1+ e^{\beta \tilde{\varepsilon}_{{\bf k}}^{f}}}
  \left(
    \frac{1}{2} +  
    \frac{1}{2}
    \frac{
      \varepsilon_{{\bf k}}-\varepsilon_{f}
    }
    {
      \sqrt{
        (\varepsilon_{{\bf k}}-\varepsilon_{f})^{2} + 4 |V_{{\bf k}}|^{2}
      }
    }
  \right)  \nonumber \\
  &&
  + \,
  \frac{1}{1+ e^{\beta \tilde{\varepsilon}_{{\bf k}}^{c}}}
  \left(
    \frac{1}{2} - 
    \frac{1}{2}
    \frac{
      \varepsilon_{{\bf k}}-\varepsilon_{f}
    }
    {
      \sqrt{
        (\varepsilon_{{\bf k}}-\varepsilon_{f})^{2} + 4 |V_{{\bf k}}|^{2}
      }
    }
  \right)  \nonumber 
\end{eqnarray}
where Eqs. \eqref{31} and \eqref{32} was used.
On the other hand, static and dynamic expectation values 
can be evaluated by applying the unitary transformations not only on the 
Hamiltonian but on all operator quantities which appear in  
expectation values. By exploiting that operator expressions under 
a trace are invariant against unitary transformations the 
$f$ electron occupation number is also given by
\begin{eqnarray}
  \label{34}
  \langle f_{{\bf k}m}^{\dagger}f_{{\bf k}m} \rangle &=&
  \langle 
    f_{{\bf k}m}^{\dagger}(\lambda) \, f_{{\bf k}m}(\lambda) 
  \rangle_{\lambda} \\
  &=& 
  \langle f_{{\bf k}m}^{\dagger}(\lambda\rightarrow 0) \, 
  f_{{\bf k}m}(\lambda\rightarrow 0)  \rangle_{(\lambda\rightarrow 0)} 
  \nonumber
\end{eqnarray}
where 
\begin{eqnarray}
  \label{35}
  f_{{\bf k}m}^{\dagger}(\lambda) &=& 
  e^{X_\lambda} \ f_{{\bf k}m}^{\dagger}\ e^{-X_\lambda} .
\end{eqnarray}
Here, $\big< ... \big >_{\lambda}$ means the expectation value formed with 
${\cal H}_{\lambda}$. One possible way to proceed is to 
derive flow equations for the $\lambda$-dependent operators 
$f_{{\bf k}m}^{\dagger}(\lambda),
f_{{\bf k}m}(\lambda)$. However, this may lead to inconsistent approximations 
which are necessary to solve the obtained flow equations. On the other hand, 
this problem can be avoided by using an appropriate ansatz for the 
$\lambda$-dependence. Therefore, for the $\lambda$-dependent operators 
$f_{{\bf k}m}^{\dagger}(\lambda)$ and for the second operator 
$c_{{\bf k}m}^{\dagger}(\lambda)$ we shall use 
\begin{eqnarray}
  \label{36}
  f_{{\bf k}m}^{\dagger}(\lambda) &=& 
  u_{{\bf k},\lambda} \, f_{{\bf k}m}^{\dagger} + 
  v_{{\bf k},\lambda} \, c_{{\bf k}m}^{\dagger} 
  \qquad \mbox{and}\\
&& \nonumber \\
  c_{{\bf k}m}^{\dagger}(\lambda) &=& 
  -v_{{\bf k},\lambda} \, f_{{\bf k}m}^{\dagger} + 
  u_{{\bf k},\lambda} \, c_{{\bf k}m}^{\dagger} 
\nonumber 
\end{eqnarray}
with $| u_{{\bf k},\lambda}|^2 + |v_{{\bf k},\lambda}|^2 =1$. This form is 
suggested by repeatedly applying the unitary transformation on  
$f_{{\bf k}m}^{\dagger}$ and  $c_{{\bf k}m}^{\dagger}$. With Eq. \eqref{36} 
the expectation value $\big <  f_{{\bf k}m}^{\dagger} \  f_{{\bf k}m} \big>$ 
can easily be evaluated. One finds  
\begin{eqnarray}
  \label{36a}
  \langle f_{{\bf k}m}^{\dagger}f_{{\bf k}m} \rangle &=& 
  \langle f_{{\bf k}m}^{\dagger}(\lambda \rightarrow 0) 
  f_{{\bf k}m}(\lambda \rightarrow 0) 
  \rangle_{(\lambda \rightarrow 0)} \\[2ex]
  &=& 
  |u_{{\bf k},(\lambda\rightarrow 0)}|^{2}
  \frac{1}{1+e^{\beta\tilde{\varepsilon}_{{\bf k}}^{f}}} + 
  |v_{{\bf k},(\lambda\rightarrow 0)}|^{2}
  \frac{1}{1+e^{\beta\tilde{\varepsilon}_{{\bf k}}^{c}}} . \nonumber
\end{eqnarray}
If one compares both results for 
$\langle f_{{\bf k}m}^{\dagger}f_{{\bf k}m} \rangle$ the prefactors 
$u_{{\bf k},(\lambda\rightarrow 0)}$ and $u_{{\bf k},(\lambda\rightarrow 0)}$ 
can be found
\begin{eqnarray}
  \label{36b}
  |u_{{\bf k},(\lambda\rightarrow 0)}|^{2} &=&
  \frac{1}{2} + 
  \frac{1}{2}
  \frac{
    \varepsilon_{{\bf k}}-\varepsilon_{f}
  }
  {
    \sqrt{
      (\varepsilon_{{\bf k}}-\varepsilon_{f})^{2} + 4 |V_{{\bf k}}|^{2}
    }
  }, \\
&& \nonumber \\
  |v_{{\bf k},(\lambda\rightarrow 0)}|^{2} &=&
  \frac{1}{2} - 
  \frac{1}{2}
  \frac{
    \varepsilon_{{\bf k}}-\varepsilon_{f}
  }
  {
    \sqrt{
      (\varepsilon_{{\bf k}}-\varepsilon_{f})^{2} + 4 |V_{{\bf k}}|^{2}
    }
  } . \nonumber \\
&& \nonumber 
\end{eqnarray}
This result agrees  with Eq.~\eqref{20}. Thus we have obtained 
the exact transformations for the  $f$ and $c$ operators. It follows that 
all static and  and dynamic quantities involving electron creation and 
annihilation operators can be calculated in the framework of the 
renormalization approach.

\section{Application to periodic Anderson model}\label{Anderson}

In the following section we will refer to the Anderson lattice
model \cite{Anderson}  which describes an interacting system of conduction 
electrons and correlated localized $4f$ electrons, arranged periodically on a 
lattice. Within a simplified version the Hamiltonian of the model can be 
written as
\begin{eqnarray}
  \label{37}
  {\cal H} &=& {\cal H}_{0} + {\cal H}_{1} , \\[2ex]
  {\cal H}_0 & = & \varepsilon_{f} \sum _{i,m} \hat{f}^{\dagger} _{im}
  \hat{f} _{im}
  + \sum _{{\bf k},m} \varepsilon_{{\bf k}} \ c^{\dagger}_{{\bf k}m} 
  c_{{\bf k}m} , \nonumber\\
  {\cal H}_1 & = & \frac{1}{\sqrt{N}} \sum _{{\bf k},i,m} V_{{\bf k}}
  \left( 
    \hat{f}^{\dagger} _{im} c_{{\bf k}m} \,
    e^{{\rm i}{\bf k}{\bf R}_i} + {\rm h.c.} 
  \right ) .
  \nonumber
\end{eqnarray}
Here, $i$ is the $4f$ site index, ${\bf k}$ is the conduction electron
wave vector, and $V_{\bf k}$ is the hybridization matrix element
between conduction and localized electrons. $\varepsilon_{f}$ and
$\varepsilon_{{\bf k}}$, both measured from the chemical potential $\mu$,
are the excitation energies for $4f$ and conduction electrons, respectively.
As a simplification, both types of electrons are assumed to
have the same angular momentum index $m$ with $\nu_{f}$ values, 
$m=1 ... \nu_{f}$.
This will make it possible to classify terms in an $1/\nu_{f}$ expansion
as is well known from the one-impurity model \cite{Hewson}. Finally, the
local Coulomb repulsion $U_f$ at the $4f$ sites has been assumed
to be infinitely large
so that localized sites can either be empty or singly occupied, i.e., the 
${\hat f}_{im}^{\dagger}$ is defined by 
\begin{eqnarray}
  \label{38}
  \hat{f}^{\dagger} _{im} = f^{\dagger} _{im} \prod_{\tilde m (\ne m)}
  (1- n^f _{i \tilde m} ) .
\end{eqnarray}

\subsection{Perturbation theory}
For the model \eqref{37} we first evaluate the effective Hamiltonian 
${\cal H}_{\lambda}$ in perturbation theory. In contrast to the general 
discussion in Sec.~\ref{pert} now we are only interested in contributions 
which renormalize  the unperturbed part ${\cal H}_{0}$ of the Hamiltonian. 
Using Eq. \eqref{13} one finds
\begin{eqnarray}
  \label{39}
  \delta {\cal H}_{0,\lambda} &=& {\cal H}_{0,\lambda} - {\cal H}_{0} \\
  &=&
  -\frac{1}{N}
  \sum_{
    \genfrac{}{}{0pt}{1}{
      \genfrac{}{}{0pt}{1}{i,{\bf k},m}{
        \left|
          \varepsilon_{{\bf k}} - \varepsilon_{f}
        \right| > \lambda
      }
    }{}
  }
  \frac{
    \left| V_{{\bf k}} \right|^{2}
  }
  {
    \varepsilon_{{\bf k}} - \varepsilon_{f}
  }
  \left[
    \hat{f}_{im}^{\dagger} c_{{\bf k}m}, c_{{\bf k}m}^{\dagger} \hat{f}_{im}
  \right] . \nonumber 
\end{eqnarray}
Note that contributions which involve different $f$ sites have been neglected 
for simplicity. In the following we restrict ourselves to the dominant order 
in $\nu_{f}$ so that Eq. \eqref{39} reads
\begin{eqnarray}
  \label{40}
  \delta {\cal H}_{0,\lambda} &=&
  \frac{1}{N}
  \sum_{
    \genfrac{}{}{0pt}{1}{
      \genfrac{}{}{0pt}{1}{i,{\bf k},m}{
        \left|
          \varepsilon_{{\bf k}} - \varepsilon_{f}
        \right| > \lambda
      }
    }{}
  }
  \frac{
    \left| V_{{\bf k}} \right|^{2}
  }
  {
    \varepsilon_{{\bf k}} - \varepsilon_{f}
  }
  \left\{
    \langle \hat{f}_{im} \hat{f}_{im}^{\dagger}  \rangle
    c_{{\bf k}m}^{\dagger}c_{{\bf k}m} + 
  \right. \\
  &&
  \left.
    + \langle c_{{\bf k}m}^{\dagger}c_{{\bf k}m} \rangle 
    \hat{f}_{im} \hat{f}_{im}^{\dagger} +
    - \langle c_{{\bf k}m}^{\dagger}c_{{\bf k}m} \rangle 
    \langle \hat{f}_{im} \hat{f}_{im}^{\dagger}  \rangle
  \right\} .  \nonumber
\end{eqnarray}
The restriction to contributions of dominant order in $\nu_{f}$ leads to 
a partial suppression of charge fluctuations on $f$ sites. 
The contributions, taken into account in Eq. \eqref{40}, describe first 
the creation of an $f$ and the annihilation of an $c$-electron followed by the 
inverse process. Therefore, the renormalization contributions in dominant 
order in $\nu_{f}$ are based on the existence of a finite probability for 
empty $f$ sites. Furthermore, in Eq.~\eqref{40} a factorization with respect 
to the full Hamiltonian ${\cal H}$ was carried out. Note that a factorization 
with respect to ${\cal H}$ instead of with ${\cal H}_{\lambda}$ is suggested 
by the following reason: 
As discussed in Sec.~\ref{pert}, ${\cal H}_{\lambda}$ results from an 
unitary transformation which is applied to the full Hamiltonian ${\cal H}$. 
Expectation values are not changed by an unitary transformation if both 
operators and the Hamiltonian are transformed. Therefore, the factorization 
with respect to the original Hamiltonian ${\cal H}$ corresponds to taking into account higher-order renormalization. 

Equation \eqref{40} can be simplified if we introduce the projection operator 
on an empty $f$ state at site $i$
\begin{eqnarray}
  \label{41}
  {\cal P}_0 (i) & = &  \prod_{\tilde m} (1- n^f_{i\tilde m})
  \,=\,
   \hat{f}_{im} \hat{f}^{\dagger}_{im} 
\\
&=& 
  1 - \sum_{\tilde{m}} \hat{f}^{\dagger}_{i\tilde{m}}\hat{f}_{i\tilde{m}}
\nonumber
\end{eqnarray}
(independent of $m$). Note that the last equation exploits the 
completeness relation on an $f$ site. Here, the $f$ occupation number 
operator,  $n_i^f =
\sum_{\tilde{m}} \hat{f}^{\dagger}_{i\tilde{m}}\hat{f}_{i\tilde{m}}$, is 
at the same time the 
projection on the singly occupied $f$ site $i$.
According to Eq. \eqref{40} the renormalization of 
the unperturbed Hamiltonian ${\cal H}_{0}$ can be written up to second order
as
\begin{eqnarray}
  \label{42}
  \delta{\cal H}_{0,\lambda} &=&   
  \delta \varepsilon_{f,\lambda} 
  \sum _{i,m} \hat{f}^{\dagger}_{im} \hat{f} _{im} + 
  \sum _{{\bf k},m} \delta \varepsilon_{{\bf k},\lambda} \, 
  c^{\dagger} _{{\bf k}m} c_{{\bf k}m} + 
  \delta E_{\lambda} ,\nonumber\\[-2ex]
  && \\
  \delta \varepsilon_{f,\lambda} &=&
  - \frac{\nu_{f}}{N}
  \sum_{{\bf k}}
  \frac{
    \left| V_{{\bf k}} \right|^{2}
  }
  {
    \varepsilon_{{\bf k}} - \varepsilon_{f}
  }
  \left\langle n_{{\bf k}m}^{c} \right\rangle\,
  \Theta\left(
    \left|
      \varepsilon_{{\bf k}} - \varepsilon_{f}
    \right| - \lambda    
  \right) , \nonumber\\
  \delta \varepsilon_{{\bf k},\lambda} &=&
  \frac{1}{N}
  \sum_{i}
  \frac{
    \left| V_{{\bf k}} \right|^{2}
  }
  {
    \varepsilon_{{\bf k}} - \varepsilon_{f}
  }
  \left\langle {\cal P}_{0}(i) \right\rangle
  \Theta\left(
    \left|
      \varepsilon_{{\bf k}} - \varepsilon_{f}
    \right| - \lambda    
  \right) , \nonumber\\
  \delta E_{\lambda} &=&
  \frac{\nu_{f}}{N}
  \sum_{i,{\bf k}}
    \frac{
    \left| V_{{\bf k}} \right|^{2}
  }
  {
    \varepsilon_{{\bf k}} - \varepsilon_{f}
  }
  \left\langle n_{{\bf k}m}^{c} \right\rangle\,
  \left(
    1 - \left\langle {\cal P}_{0}(i) \right\rangle
  \right)\nonumber\\
  &&
  \times\Theta\left(
    \left|
      \varepsilon_{{\bf k}} - \varepsilon_{f}
    \right| - \lambda    
  \right), \nonumber
\end{eqnarray}
where   $\delta \varepsilon_{{\bf k},\lambda}$ and 
$\delta \varepsilon_{f,\lambda}$ lead to the renormalization of the 
one-particle energies $\varepsilon_{{\bf k},\lambda}$ and 
$\varepsilon_{f,\lambda}$. $\delta E_{\lambda}$ is an additional energy shift.

\subsection{Renormalization equations}
The above result in perturbation can be used to 
derive flow equations for the periodic Anderson model along the line discussed 
in Sec.~\ref{Ren}. Let us start by formally writing down the effective 
Hamiltonian ${\cal H}_{\lambda}$ after all excitations with energy differences 
larger than $\lambda$ have been eliminated. ${\cal H}_{\lambda}$ should have 
the following form
\begin{eqnarray}
  \label{43}
  {\cal H}_{\lambda} &=& 
  {\cal H}_{0,\lambda} + {\cal H}_{1,\lambda} , \\[2ex]
  {\cal H}_{0,\lambda} &=& 
  \varepsilon_{f,\lambda}
  \sum_{i,m}
  \hat{f}^{\dagger}_{im} \hat{f} _{im} +
  \sum_{{\bf k},m} 
  \varepsilon_{{\bf k},\lambda} \, c^{\dagger} _{{\bf k}m} c_{{\bf k}m} + 
  E_{\lambda} , \nonumber\\[2ex]
  {\cal H}_{1,\lambda} &=& {\bf P}_{\lambda}{\cal H}_{1} \nonumber \\
  &=&
  \frac{1}{\sqrt{N}}
  \sum_{
    \genfrac{}{}{0pt}{1}{
      \genfrac{}{}{0pt}{1}{i,{\bf k},m}{
        \left|
          \varepsilon_{{\bf k},\lambda} - \varepsilon_{f,\lambda}
        \right| \leq \lambda
      }
    }{}
  }
  V_{{\bf k}}
  \left(
    \hat{f}^{\dagger} _{im} c_{{\bf k}m} \,
    e^{{\rm i}{\bf k}{\bf R}_i} + {\rm h.c.}
  \right) . \nonumber
\end{eqnarray}
In the following we are only interested in the renormalization of the diagonal 
part of the Hamiltonian. Consequently, all additional terms which would appear 
due to Eqs. \eqref{16} and \eqref{17} will be neglected in the following for 
simplicity. Due to renormalization processes the one-particle energies 
$\varepsilon_{f, \lambda}$, $\varepsilon_{{\bf k},\lambda}$, 
as well as the energy shift $E_{\lambda}$ now depend on the energy cutoff 
$\lambda$. The initial conditions are those from the original model 
\begin{eqnarray}
  \label{44}
  \varepsilon_{f, (\lambda =  \Lambda)} &=& 
  \varepsilon_f, \quad 
  \varepsilon_{{\bf k}, (\lambda = \Lambda)} \,=\,
  \varepsilon_{{\bf k}}, \quad
  E_{(\lambda = \Lambda)} \,=\, 0 .
\end{eqnarray}
In the following step let us evaluate a new effective Hamiltonian which is 
obtained after a further elimination of excitations within a small energy 
shell between $\lambda-\Delta\lambda$ and $\lambda$ has been done. The new 
Hamiltonian ${\cal H}_{(\lambda-\Delta\lambda)}$ is obtained from 
Eq.~\eqref{42} and possesses the following renormalized parameters
\begin{eqnarray}
  \label{45}
  \varepsilon_{f,(\lambda-\Delta\lambda)} &=& \varepsilon_{f,\lambda}
  - \frac{\nu_{f}}{N}
  {\sum_{{\bf k}}}^{\prime}
  \frac{
    \left| V_{{\bf k}} \right|^{2}
  }
  {
    \varepsilon_{{\bf k},\lambda} - \varepsilon_{f,\lambda}
  }
  \left\langle n_{{\bf k}m}^{c} \right\rangle , \\
  \varepsilon_{{\bf k},(\lambda-\Delta\lambda)} &=&
  \varepsilon_{{\bf k},\lambda} + 
  \frac{1}{N}
  {\sum_{i}}^{\prime}
  \frac{
    \left| V_{{\bf k}} \right|^{2}
  }
  {
    \varepsilon_{{\bf k},\lambda} - \varepsilon_{f,\lambda}
  }
  \left\langle {\cal P}_{0}(i) \right\rangle , \nonumber\\
  E_{(\lambda-\Delta\lambda)} &=&
  E_{\lambda}+ 
  \frac{\nu_{f}}{N}
  {\sum_{i,{\bf k}}}^{\prime}
    \frac{
    \left| V_{{\bf k}} \right|^{2}
  }
  {
    \varepsilon_{{\bf k},\lambda} - \varepsilon_{f,\lambda}
  }
  \left\langle n_{{\bf k}m}^{c} \right\rangle\, \nonumber\\
  &&
  \times \left(
    1 - \left\langle {\cal P}_{0}(i) \right\rangle
  \right) . \nonumber  
\end{eqnarray}
Here, the prime $\prime$ above the sums indicates that the restriction 
$(\lambda-\Delta\lambda) < 
  |\varepsilon_{{\bf k},\lambda}-\varepsilon_{f,\lambda}| \leq 
  \lambda$
has to be fulfilled.

Finally, we would like to perform the stepwise transformation continuously. 
By taking the limit $\Delta\lambda\rightarrow 0$, we obtain the following 
flow equations for the 
parameter $\varepsilon_{f,\lambda}$, $\varepsilon_{{\bf k},\lambda}$, and 
$E_{\lambda}$ 
\begin{eqnarray}
  \label{46}
  \frac{d \varepsilon _{f, \lambda}}{d \lambda} &=& 
  \frac{\nu_{f}}{N}
  \sum_{{\bf k}}
  \frac{
    \left| V_k \right|^2
  }{
    \varepsilon_{{\bf k},\lambda}-\varepsilon _{f,\lambda}
  }
  \left\langle n_{{\bf k}m}^{c} \right\rangle
  \delta \left(
    \lambda - 
    \left|
      \varepsilon _{{\bf k}, \lambda} - \varepsilon _{f, \lambda}
    \right|
  \right) , \nonumber\\[-2ex]
  && \\
  \label{47}
  \frac{d \varepsilon _{{\bf k},\lambda}}{d \lambda} &=& 
  - \frac{
    \left| V_k \right|^2
  }{
    \varepsilon_{{\bf k},\lambda} - \varepsilon_{f,\lambda}
  }
  \left\langle {\cal P}_{0}(i) \right\rangle 
  \delta \left(
    \lambda - 
    \left| \varepsilon _{{\bf k}, \lambda} - \varepsilon _{f,\lambda} \right|
  \right) , \\[2ex]
  \label{48}
  \frac{d E_{\lambda}}{d \lambda } &=& 
  - \nu_{f} 
  \sum_{{\bf k}} 
  \frac{
    \left| V(k) \right|^2
  }{
    \varepsilon _{{\bf k}, \lambda} - \varepsilon _{f, \lambda}
  }
  \left\langle n_{{\bf k}m}^{c} \right\rangle 
  \left(
    1 - \left\langle {\cal P}_{0}(i) \right\rangle
  \right) \\
  &&
  \times\delta \left(
    \lambda - 
    \left| \varepsilon _{{\bf k}, \lambda} - \varepsilon _{f, \lambda} \right|
  \right)) \nonumber\\
  &=& 
  - N  
  \left(
    \left\langle 1 - {\cal P}_{0}(i) \right\rangle
  \right)
  \frac{ d\varepsilon_{f,\lambda} }{d\lambda} . \nonumber
\end{eqnarray}
Here, due to the ${\bf k}$ summation in Eq.~\eqref{46} it was assumed that the 
renormalization of the one-particle energies $\varepsilon_{f,\lambda}$ and 
$\varepsilon_{{\bf k},\lambda}$ is continuous in contrast to the case of 
Sec.~\ref{Fano}. The first two equations are coupled nonlinear differential 
equations for the parameters $\varepsilon_{f,\lambda}$ and 
$\varepsilon_{{\bf k},\lambda}$. Here, the $\delta$-function guarantees that 
only excitation on the energy shell contribute.  

To solve Eqs. \eqref{46} and \eqref{47} first note that due to 
translational invariance the expectation value $\left\langle {\cal P}_{0}(i) \right\rangle$ 
is independent of site index $i$. Therefore, a simple relation between 
$\varepsilon_{f,\lambda}$ and $ \varepsilon_{{\bf k},\lambda}$ can be derived 
by inserting \eqref{47} into \eqref{46}
\begin{eqnarray}
  \label{49}
  \left\langle {\cal P}_{0}(i) \right\rangle 
  \frac{d\varepsilon_{f,\lambda}}{d\lambda} &=&
  - \frac{\nu_{f}}{N}
  \sum_{{\bf k}} 
  \left\langle n_{{\bf k}m}^{c} \right\rangle
  \frac{d \varepsilon _{{\bf k},\lambda}}{d \lambda} .
\end{eqnarray}
Its integration between the lower cutoff 
$\lambda\rightarrow 0$ and the cutoff $\Lambda$ of the original model leads 
to
\begin{eqnarray}
  \label{50}
  \left\langle {\cal P}_{0}(i) \right\rangle
  \left(
    \varepsilon_{f} - \tilde{\varepsilon}_{f}
  \right)
  &=&
  - \frac{\nu_{f}}{N}
  \sum_{{\bf k}} 
  \left\langle n_{{\bf k}m}^{c} \right\rangle
  \left(
    \varepsilon_{{\bf k}} - \tilde{\varepsilon}_{{\bf k}}
  \right)
\end{eqnarray}
where we have defined 
$ \tilde{\varepsilon}_{f}= \varepsilon_{f,(\lambda\rightarrow 0)}$ and  
$ \tilde{\varepsilon}_k=\varepsilon _{{\bf k},(\lambda\rightarrow 0)} $.
Note that as before the expectation values in Eq. \eqref{50} are 
assumed to be independent of $\lambda$  since they were defined with the 
full Hamiltonian. Eq.\eqref{50} is the first equation which 
relates $\tilde{\varepsilon}_{f}$ and $\tilde{\varepsilon}_k$. 
To find a second equation let us integrate 
Eq.~\eqref{47} between the lower cutoff $(\lambda\rightarrow 0)$ and the 
cutoff $\Lambda$ of the original model. One finds
\begin{eqnarray}
  \label{51}
  \tilde{\varepsilon}_{{\bf k}} &=& 
  \varepsilon _k +
  \int^{\Lambda}_{(\lambda\rightarrow 0)} d \lambda^{\prime}
  \frac{
    \left| V_k \right|^2 
    \left\langle {\cal P}_{0}(i) \right\rangle 
  }{
    \varepsilon_{{\bf k},\lambda^{\prime}} - \varepsilon _{f, \lambda^{\prime}}
  }\\
  && 
  \times\delta\left(
    \lambda^{\prime} - 
    \left|
      \varepsilon _{{\bf k}, \lambda^{\prime}} - 
      \varepsilon _{f, \lambda^{\prime}}
    \right|
  \right)\nonumber
\end{eqnarray}
where Eq. \eqref{44} was used. Since the flow of the one-particle energies 
$\varepsilon_{{\bf k},\lambda}$ and $\varepsilon_{f,\lambda}$ was assumed to 
be continous, the main contribution to the integral in Eq. \eqref{51} should 
originate from small values of the denominators 
$
  {\varepsilon _{{\bf k},\lambda^{\prime}} - 
  \varepsilon_{f,\lambda^{\prime}}}
$, 
i.e., from small values of $\lambda^{\prime}$. Therefore, a reasonable 
approximation is to replace the quantity   
$
  {\varepsilon _{{\bf k},\lambda^{\prime}} - 
  \varepsilon _{f,\lambda^{\prime}}}
$ 
in Eq. \eqref{51} by its value at the lower cutoff 
$\lambda^{\prime} \approx\lambda\rightarrow 0$. Thus we obtain
\begin{eqnarray}
  \label{52}
  \tilde{\varepsilon}_{{\bf k}} &=& 
  \varepsilon _{{\bf k}} +
  \frac{
    \left| V_k \right|^2 
    \left\langle {\cal P}_{0}(i) \right\rangle 
  }{
    \tilde{\varepsilon}_{{\bf k}} - \tilde{\varepsilon}_{f}
  } . 
\end{eqnarray}
Note that additional contributions of higher order in $|V_{\bf k}|^2$
have been neglected. 

Equations \eqref{52} and \eqref{50} fix the one-particle energies. The 
Hamiltonian ${\cal H}_{(\lambda \rightarrow 0)}$ reads
\begin{eqnarray}
  \label{54}
  {\cal H}_{(\lambda\rightarrow 0)} &=& 
  \tilde{\varepsilon}_{f} \sum _{i,m}
  \hat f_{im} ^{\dagger} \hat f_{im} +
  \sum _{{\bf k},m} 
  \tilde{\varepsilon}_{{\bf k}} \, c^{\dagger}_{{\bf k}m} c_{{\bf k}m} + 
  \tilde{E} 
\end{eqnarray}
where 
\begin{eqnarray}
  \label{55}
  \tilde{\varepsilon}_{f} &=& 
  \varepsilon_{f} - \frac{\nu_{f}}{N} 
  \sum_{{\bf k}} 
  \frac{
    \left| V_k \right|^2 
    \left\langle  n_{{\bf k}m}^{c} \right\rangle
  }{
    \tilde{\varepsilon}_{{\bf k}} - \tilde{\varepsilon}_{f}
  }, \\[2ex]
  \label{56}
  \tilde{\varepsilon}_{{\bf k}} &=& 
  \varepsilon _{{\bf k}} +
  \frac{
    \left| V_k \right|^2 
    \left\langle {\cal P}_{0}(i) \right\rangle 
  }{
    \tilde{\varepsilon}_{{\bf k}} - \tilde{\varepsilon}_{f}
  },\\[2ex]
  \label{57}
  \tilde{E} &=& 
  N \left(
    1 - \left\langle {\cal P}_{0}(i) \right\rangle
  \right)
  \left(
    \varepsilon_{f} - \tilde{\varepsilon}_{f}
  \right).
\end{eqnarray}
As before, the final model ${\cal H}_{(\lambda \rightarrow 0)}$ consists of 
noninteracting $f$ and conduction electrons with renormalized one-particle 
energies $\tilde{\varepsilon}_{f}$ and $\tilde{\varepsilon}_{\bf k}$. The 
expectation values 
$ 
  \left\langle {\cal P}_0(i) \right\rangle = 
  1 - \langle n_{i}^{f} \rangle
$ 
and 
$\left\langle n_{{\bf k}m}^{c} \right\rangle$ in Eqs. \eqref{55} and 
\eqref{57} can again be evaluated from the free energy (for details see 
Appendix B).

Let us solve Eq.~\eqref{56} for $\tilde{\varepsilon}_{{\bf k}}$. We find two 
solutions 
\begin{eqnarray}
  \label{59}
  \tilde{\varepsilon}_{{\bf k} (1,2)} &=& 
  \frac{1}{2}\left\{
    \varepsilon_{{\bf k}} + \tilde{\varepsilon}_{f} 
    \mp
    \sqrt{
      \left( \varepsilon_{{\bf k}} - \tilde{\varepsilon}_{f} \right )^2
      + 
      4 \left| V_{\bf k}\right|^2 \left\langle {\cal P}_{0}(i) \right\rangle
    }
  \right\} \nonumber\\ 
  &=& 
  \frac{1}{2}\left\{ 
    \varepsilon_{{\bf k}} + \tilde{\varepsilon}_{f} \mp W_{{\bf k}} 
  \right\}
\end{eqnarray}
which agree with the well--known two heavy quasiparticle bands
known from the slave-boson formalism \cite{Fulde}.  
In the following let us  assume that the renormalized $f$ level 
$\tilde{\varepsilon}_f$ 
lies slightly above the chemical potential $\mu$ whereas the lower 
quasiparticle band $\tilde{\varepsilon}_{{\bf k} (1)}$ is intersected by $\mu$. 
As is easily seen our result is somewhat different from that obtained in 
slave-boson mean-field theory. The renormalized one-particle excitation 
$\tilde{\varepsilon}_{\bf k}$ in Eq. \eqref{59} agrees with the lower heavy 
quasiparticle band of the slave-boson theory 
$\tilde{\varepsilon}_{\bf k} = \tilde{\varepsilon}_{{\bf k}(1)}$. 
However, the renormalized $f$ electron excitation $\tilde{\varepsilon}_f$ is
different from the upper heavy quasiparticle band. Instead we have found a 
dispersion less excitation which is located slightly above the Fermi level
(see appendix B). Note that for temperature $T=0$ both approaches lead to same result for 
thermodynamic quantities since at zero temperature only the lower quasiparticle 
part is filled up to the Fermi level. However, static properties for $T \ne 0$ as 
well as dynamic correlation function turn out to be different within the 
two approaches.

\subsection{Transformed operators and electron Green's function}
To evaluate dynamic and different static quantities we have again to apply 
the transformation on operator quantities inside the expectation values. 
For the one-particle Green's functions we need 
\begin{eqnarray}
\label{60a}
c_{{\bf k}m}^{\dagger}(\lambda) &=& e^{X_{\lambda}} \ c_{{\bf k}m}^{\dagger} \  e^{-X_{\lambda}}
\end{eqnarray}
and also $f_{im}^{\dagger}(\lambda)$.  In first order in the hybridization 
$V_{\bf k}$ one finds for the conduction electron creation operator 
\begin{eqnarray}
  \label{60}
  c_{{\bf k}m}^{\dagger}(\lambda) &\approx&
  c_{{\bf k}m}^{\dagger} + 
  \left[
    \frac{1}{L_{0}} 
    \left( {\bf Q}_{\lambda} {\cal H}_{1}\right) , c_{{\bf k}m}^{\dagger}
  \right]\\
  &=&
  c_{{\bf k}m}^{\dagger} + 
  \frac{V_{{\bf k}}}{\varepsilon_{f}-\varepsilon_{{\bf k}}} 
  \hat{f}^{\dagger}_{{\bf k}m} \nonumber
\end{eqnarray}
where $\hat{f}^{\dagger}_{{\bf k}m}$ is the Fourier transform of the 
$\hat{f}^{\dagger}_{im}$. This result suggests to make the following ansatz 
\begin{eqnarray}
  \label{61}
  c_{{\bf k}m}^{\dagger}(\lambda) &=& 
  u_{{\bf k}, \lambda} \, c_{{\bf k}m}^{\dagger} + 
  v_{{\bf k}, \lambda} \, \hat{f}^{\dagger}_{{\bf k}m}
\end{eqnarray}
which should be suited to obtain a good approximation for 
$c_{{\bf k}m}^{\dagger}(\lambda)$. Because  $c_{{\bf k}m}^{\dagger}(\lambda), c_{{\bf k}m}(\lambda)$
have to be fulfill the Fermi anticummutator relations, one concludes that  
\begin{eqnarray}
  \label{62}
  1 &=& 
  \left| u_{{\bf k}, \lambda} \right|^{2} + 
  \left| v_{{\bf k}, \lambda} \right|^{2} \left\langle {\cal P}_{0}(i)\right\rangle
\end{eqnarray}
holds for all values of ${\bf k}$. Thereby, the approximation
\begin{eqnarray}
  \label{63}
  \left[
    \hat{f}^{\dagger}_{{\bf k}m}, \hat{f}_{{\bf k}m}
  \right]_{+} 
  &=& 
  \left\langle {\cal P}_{0}(i) \right\rangle
\end{eqnarray}
was used. By using Eq. \eqref{61}, the conduction electron occupation 
probability reads 
\begin{eqnarray}
  \label{64}
  && \\[-3ex]
  \left\langle n_{{\bf k}m}^{c} \right\rangle &=&
  \langle c_{{\bf k}m}^{\dagger}c_{{\bf k}m} \rangle \,=\,
  \left\langle 
    c_{{\bf k}m}^{\dagger}(\lambda\rightarrow 0)
    c_{{\bf k}m}(\lambda\rightarrow 0)
  \right\rangle_{(\lambda\rightarrow 0)} \nonumber\\
  &=& 
  \left| u_{{\bf k},(\lambda\rightarrow 0)} \right|^{2}
  \frac{1}{1 + e^{\beta\tilde{\varepsilon}_{{\bf k}}}} + 
  \left| v_{{\bf k},(\lambda\rightarrow 0)} \right|^{2}
  \frac{1}{\nu_{f} + e^{\beta\tilde{\varepsilon}_{f}}} \nonumber
\end{eqnarray}
where $\langle ... \rangle_{(\lambda\rightarrow 0)}$ again denotes the 
expectation value formed with the renormalized Hamiltonian 
${\cal H}_{(\lambda\rightarrow 0)}$. Note that 
\begin{eqnarray}
  \label{64a}
\big< {\hat f}_{im}^{\dagger} {\hat f}_{im} \big >_{(\lambda \rightarrow 0)} &=&
\frac{1}{\nu_f + e^{\beta \tilde{\varepsilon}_f}}
\end{eqnarray}
was used in Eq. \eqref{64} which differs from the Fermi distribution due to 
definition \eqref{38} for the modified creation and annihilation operators 
${\hat f}_{im}^{\dagger}$ and ${\hat f}_{im}$. 

On the other hand, the expectation value 
$\langle c_{{\bf k}m}^{\dagger}c_{{\bf k}m} \rangle$ can also be calculated by 
functional derivative from  the free energy (for details see Appendix B). 
For zero temperature one obtains
\begin{eqnarray}
  \label{65}
  \left\langle n_{{\bf k}m}^{c} \right\rangle &=&
  \frac{1}{2}
  \left(
    1 - \frac{ \varepsilon_{{\bf k}}-\tilde{\varepsilon}_{f} }{ W_{{\bf k}} }
  \right)
  \Theta(k_{F}-k)
\end{eqnarray}
where $W_{\bf k}$ was defined by Eq. \eqref{59}. By comparing Eq. \eqref{64} 
with Eq. \eqref{65} one finds ($k \leq k_{F}, \beta \rightarrow \infty$)
\begin{eqnarray}
  \label{66}
  \left| u_{{\bf k},(\lambda\rightarrow 0)} \right|^{2}
  &=&
  \frac{1}{2}
  \left(
    1 - \frac{ \varepsilon_{{\bf k}}-\tilde{\varepsilon}_{f} }{ W_{{\bf k}} }
  \right)
\end{eqnarray}
and
\begin{eqnarray}
  \label{67}
  \left| v_{{\bf k},(\lambda\rightarrow 0)} \right|^{2}
  &=&
  \frac{1}{2\left\langle {\cal P}_{0}(i) \right\rangle}
  \left(
    1 + \frac{ \varepsilon_{{\bf k}}-\tilde{\varepsilon}_{f} }{ W_{{\bf k}} }
  \right)
\end{eqnarray}
where condition \eqref{62} was used.
For the transformed $f$ electron creation operator one finds  
\begin{eqnarray}
  \label{68}
  \hat{f}^{\dagger}_{{\bf k}m}(\lambda) &=&
  -\left\langle {\cal P}_{0}(i)\right\rangle 
  v_{{\bf k},\lambda} c_{{\bf k}m}^{\dagger} +
  u_{{\bf k},\lambda} \hat{f}^{\dagger}_{{\bf k}m}
\end{eqnarray}
where once again Eq. \eqref{63} was used. We are now able to determine the 
$f$ electron Green's function
\end{multicols}
\widetext
\begin{eqnarray}
  \label{69}
  G_{{\bf k}m}(\omega) &=&
  \left\langle
    \left[
      \hat{f}_{{\bf k}m}, \frac{1}{{\bf L}-(\omega + i \eta)}
      \hat{f}_{{\bf k}m}^{\dagger}
    \right]_{+}
  \right\rangle \,=\,
  \left\langle
    \left[
      \hat{f}_{{\bf k}m}(\lambda\rightarrow 0),
      \frac{1}{{\bf L}_{(\lambda\rightarrow 0)}-(\omega + i \eta)}
      \hat{f}_{{\bf k}m}^{\dagger}(\lambda\rightarrow 0)
    \right]_{+}
  \right\rangle_{(\lambda\rightarrow 0)}
\end{eqnarray}
\begin{multicols}{2}
\narrowtext
where new Liouville operators ${\bf L}$ and  ${\bf L}_{\lambda}$ with respect to 
${\cal H}$ and ${\cal H}_{\lambda}$ were introduced. For the imaginary
part of the $f$  electron Green's function we obtain
\begin{eqnarray}
  \label{70}
  {\rm Im}\, G_{{\bf k}m}(\omega) &=&
  \left\langle {\cal P}_{0}(i)\right\rangle
  \left\{
    \left\langle {\cal P}_{0}(i)\right\rangle
    \left| v_{{\bf k},(\lambda\rightarrow 0)} \right|^{2}
    \delta\left(
      \omega - \tilde{\varepsilon}_{{\bf k}}
    \right) 
  \right.\nonumber\\
  && 
  \left.
    + \left| u_{{\bf k},(\lambda\rightarrow 0)} \right|^{2}
    \delta\left(
      \omega - \tilde{\varepsilon}_{f}
    \right)
  \right\}
\end{eqnarray}
where again $k\leq k_{F}$ has to be fulfilled. As is seen from Eqs. \eqref{66} 
and \eqref{67} for small $k$ values the peak at $\tilde{\varepsilon}_{f}$ 
dominates the imaginary part of the $f$-electron Green's function. The pole 
at $\tilde{\varepsilon}_{{\bf k}}$ becomes important for $k$ values near the Fermi 
momentum $k_{F}$. 

\subsection{Discussion of the results}
Let us compare our  results for the periodic 
Anderson model with those obtained from the slave-boson mean 
field theory. For simplicity we will restrict ourselves to the $T=0$ case.
As already mentioned above, the renormalized energy 
$\tilde{\varepsilon}_{{\bf k}}$ corresponds to the lower quasiparticle bands 
from the slave-boson mean field treatment. Furthermore, the renormalized  
Hamiltonian ${\cal H}_{(\lambda \rightarrow 0)}$ leads to 
same $f$ state occupation as the slave-boson theory 
(for details see Appendix B). Moreover, the typical Kondo peak at the Fermi 
level is found in the imaginary part of the $f$ electron Green's function 
\eqref{70}. Therefore, the renormalization approach for 
the periodic Anderson model leads at $T=0$ to the same results as 
the slave-boson mean field theory. On the other hand, some differences 
occur. The most important one is the additional quasi-particle band at the 
renormalized $f$ level $\tilde{\varepsilon}_{f}$ which displaces the upper 
heavy quasiparticle band from the slave-boson mean-field treatment. Note 
however that in the present approach we have neglected all processes which 
include different $f$ sites. A detailed discussion will be given in a 
forthcoming paper \cite{Sommer}.

\section{Conclusion}\label{Conc}

In summary, in this paper we have presented a renormalization approach to 
many-particle Hamiltonians. First, 
the elimination of high energy transition operators larger than 
an energy cutoff $\lambda$ leads to a transformed 
Hamiltonian ${\cal H}_{\lambda} $which is  band diagonal 
with respect to the eigenbasis of 
its unperturbed part.   
The approach, which is based on perturbation theory, can be extended 
to establish a renormalization approach by continuously shifting the energy
cutoff $\lambda$ to lower and lower values. In this way 
flow equations for the Hamiltonian are obtained. 
The present  approach 
has some similarities with recently introduced renormalization 
method by Wegner \cite{Wegner1} and by Glatzek and Wilson 
\cite{Wilson1,Wilson2}. However, there are also some substantial differences: 
The present method starts from unitary transformations which are directly  
applied to the Hamiltonian. In contrast, the previous approach is 
formulated in matrix notation. It starts from flow equations  in differential form 
for the Hamiltonian as well as for the generator of the 
unitary transformation. The present flow equations for the 
Hamiltonian can be formulated independent from a concrete 
representation.  
For demonstration the method was first applied to the exactly  
solvable Fano-Anderson model and to  the Anderson-lattice model.
For the first model the exact results were found. For the second model some 
approximations had to be done. For this case the 
well-known quasiparticle behavior of heavy fermions was rederived. In 
particular, it was shown that there is an almost complete agreement with the 
results from the slaved-boson mean-field theory \cite{Fulde}.

\section*{Acknowledgments}
We would like to acknowledge fruitful discussions with F.~Anders, W.~Brenig, 
P.~Fulde, N.~Grewe, C. K\"{u}hnert, K.~Meyer, A.~Muramatsu, and S.~Sykora. 
This work was supported by DFG through the research program SFB 463.

\begin{appendix}

\section{Flow equation in matrix notation} 
Let us start from the eigenvalue problem of the unperturbed 
Hamiltonian ${\cal H}_{0,\lambda}$ after all transitions with energies 
larger than $\lambda$ have been integrated out
\begin{eqnarray}
  \label{A1}
  {\cal H}_{0,\lambda}\, \left| n(\lambda) \right\rangle &=& 
  E_{n,\lambda}^{0} \, \left| n(\lambda) \right\rangle .
\end{eqnarray}
The eigenvectors of ${\cal H}_{0, \lambda}$ may depend on $\lambda$.
We consider two degenerate eigenstates 
$|n(\lambda) \big >, |\tilde{n}(\lambda) \big > $ 
belonging to the same energy  $E_{n,\lambda}^0 = E_{\tilde{n},\lambda}^0$.
Let us assume that the eigenstates have been orthogonalized, i.e.,   
$\big< n(\lambda)|  \tilde{n}(\lambda) \big > = \delta_{n, \tilde{n}} $.  
We multiply Eq. \eqref{16} from the left with $\big< n(\lambda)|$
and from the right with $|\tilde{n}(\lambda) \big >$.  
For the renormalization of ${\cal H}_{0, \lambda}$ due to the elimination
of excitations within the energy shell between $\lambda -\Delta\lambda$ 
and $\lambda$ one finds
\begin{eqnarray}
  \label{A2}
  \lefteqn{
    \left\langle 
      n(\lambda) | ( {\cal H}_{0,(\lambda -\Delta\lambda)} - 
      {\cal H}_{0,\lambda} ) |\tilde{n}(\lambda) 
    \right\rangle
    \,=\,
  }\\
  &=&  
  - \mbox{\hspace{-4ex}}\sum_{
    \genfrac{}{}{0pt}{1}{
      \genfrac{}{}{0pt}{1}{m}{
        \lambda-\Delta\lambda < 
        \left|
          E_{n,\lambda}^{0} - E_{m,\lambda}^{0}
        \right|
        \leq \lambda
      }
    }{}
  }\mbox{\hspace{-4ex}}
\frac{
\big<n(\lambda)| {\cal H}_{1,\lambda} |m(\lambda)\big > 
\big<m(\lambda)| {\cal H}_{1,\lambda} |\tilde{n}(\lambda)\big >} {
  E_{m,\lambda}^0 - E_{n,\lambda}^0 } . \nonumber
\end{eqnarray}
Note that the energy constraint 
$\lambda-\Delta\lambda < |E_{n,\lambda}^0 -E_{m,\lambda}^0| \le \lambda $ 
may also be written as difference of two $\Theta$ functions
\begin{eqnarray}
\label{A3}
\lefteqn{
  \sum_{\lambda -\Delta\lambda < |E_{n,\lambda}^0 -E_{m,\lambda}^0| \le
  \lambda } \cdots \ \ \ 
  \,=\,
} \qquad \qquad && \\
&=& \sum \cdots \ \ 
\left\{
  \Theta[|E_{n,\lambda}^0 -E_{m,\lambda}^0| -(\lambda - \Delta\lambda)]
\right. \nonumber\\
&&
\left.
  -\Theta(|E_{n,\lambda}^0 -E_{m,\lambda}^0| -\lambda )
\right\} \nonumber
\end{eqnarray}
so that on the rhs of Eq. \eqref{A2} the limit $\Delta\lambda \rightarrow 0$ 
can be easily performed.
For the lhs it is easy to verify that the limit $\Delta\lambda \rightarrow 0$
leads to the derivative of $E_{n,\lambda}^0$ with respect to 
$\lambda$. Here, 
the orthogonality of the vectors $|n(\lambda) \big>, 
|\tilde{n}(\lambda) \big>$ has to be used. Thus, one  finds 
\begin{eqnarray}
\label{A4}
- \frac{dE_{n,\lambda}^0}{d\lambda} &=&
- \sum_{m} \frac{
\big<n(\lambda)| {\cal H}_{1,\lambda} |m(\lambda)\big > 
\big<m(\lambda)| {\cal H}_{1,\lambda} |\tilde{n}(\lambda)\big >}
{E_{m,\lambda}^0 - E_{n,\lambda}^0} \nonumber\\
&&
\times
\frac{d\Theta(|E_{n,\lambda}^0 -E_{m,\lambda}^0| -\lambda)} {d\lambda} .
\end{eqnarray}
In Eq. \eqref{A4} the $\lambda$ derivative of the $\Theta$ function 
acts on  the explicit term linear in $\lambda$  
and on the $\lambda$-dependent energies $E_{n,\lambda}^0$ and 
$E_{m,\lambda}^0$. It is obvious that the latter dependence 
leads to higher order contributions in ${\cal H}_{1,\lambda}$. 
Therefore, by restricting one-self to the lowest non vanishing order 
in ${\cal H}_{1,\lambda}$ one finds 
\begin{eqnarray}
\label{A5}
\frac{dE_{n,\lambda}^0}{d\lambda} &=&
- \sum_{m} 
\big<n(\lambda)| {\cal H}_{1,\lambda} |m(\lambda)\big > 
\big<m(\lambda)| {\cal H}_{1,\lambda} |\tilde{n}(\lambda)\big > \nonumber \\
&&
\times
\frac{1}{  E_{m,\lambda}^0 - E_{n,\lambda}^0}\
\delta(|E_{n,\lambda}^0 -E_{m,\lambda}^0| -\lambda) . 
\end{eqnarray}   
Note that flow equations for the $\lambda$ dependent matrix elements 
of the interaction ${\cal H}_{1,\lambda}$ can be derived in a similar way.

\section{Calculation of the occupation probabilities}

In this appendix we derive formal expressions for the particle 
number expectation values for the periodic Anderson model. 
Thereby, we restrict ourselves to zero temperature.
Let us  assume that the renormalized $f$ level 
$\tilde{\varepsilon}_f$ lies slightly above the chemical potential $\mu$ 
whereas the lower quasiparticle band $\tilde{\varepsilon}_{{\bf k} (1)}$ is 
intersected by $\mu$. Thus for $T=0$,  according to  Eqs. \eqref{54} and 
\eqref{57} the ground-state energy  $E_g = F(T \rightarrow 0)$ is given by
\begin{eqnarray}
  \label{B1}
  E_g & = & 
  \tilde{E} + \nu_{f} \sum _{k\leq k_F} \tilde{\varepsilon}_{{\bf k}} \\
  &=&
  N \left(
    1 - \left\langle {\cal P}_{0}(i) \right\rangle
  \right)
  \left(
    \varepsilon_{f} - \tilde{\varepsilon}_{f}
  \right) + 
  \nu_{f} \sum_{k\leq k_F} \tilde{\varepsilon}_{{\bf k}} \nonumber
\end{eqnarray}
where $k_{F}$ denotes the Fermi momentum and  
$\tilde{\varepsilon}_{{\bf k} (1)}=
\tilde{\varepsilon}_{\bf k}$. 
The ground state energy can be used 
to calculate the zero-temperature occupation probabilities.  The functional 
derivative of Eq.~\eqref{54} leads to 
\begin{eqnarray}
  \label{B2}
  \langle n_{i}^{f} \rangle &=&
  \frac{1}{N} \frac{\partial E_{g}}{\partial \varepsilon_{f}} \,=\,
  \frac{1}{N} \frac{\partial \tilde{E}}{\partial \varepsilon_{f}} +
  \frac{\nu_{f}}{N} \sum_{k\leq k_{F}}
  \frac{\partial \tilde{\varepsilon}_{{\bf k}}}{\partial \varepsilon_{f}} 
  \quad\mbox{and} \\
  \label{B3}
  \left\langle n_{{\bf k}m}^{c} \right\rangle &=&
  \frac{1}{\nu_{f}} \frac{\partial E_{g}}{\partial \varepsilon_{{\bf k}}} \,=\,
  \frac{1}{\nu_{f}}\frac{\partial \tilde{E}}{\partial \varepsilon_{{\bf k}}} +
  \sum_{k^{\prime}\leq k_{F}}
  \frac{\partial \tilde{\varepsilon}_{{\bf k}^{\prime}}}
  {\partial \varepsilon_{{\bf k}}}
\end{eqnarray}
where due to Eq.~\eqref{57} 
\begin{eqnarray}
  \label{B4}
  \frac{\partial \tilde{E}}{\partial \varepsilon_{f}} &=&
  N \langle n_{i}^{f} \rangle
  \left(
    1 - \frac{\partial \tilde{\varepsilon}_{f}}{\partial \varepsilon_{f}}
  \right) +
  N \left(
    \varepsilon_{f} - \tilde{\varepsilon}_{f}
  \right)
  \frac{\partial \langle n_{i}^{f} \rangle }
{\partial  \varepsilon_{f} }.
\end{eqnarray}
Furthermore, by using Eq. \eqref{59} one obtains
\begin{eqnarray}
  \label{B5}
  \frac{ \partial \tilde{\varepsilon}_{{\bf k}} }{ \partial \varepsilon_{f} } &=&
  \frac{1}{2}
  \left(
    1 + \frac{ \varepsilon_{{\bf k}} - \tilde{\varepsilon}_{f} }{W_{{\bf k}}}
  \right)
  \frac{\partial  \tilde{\varepsilon}_{f} }{\partial \varepsilon_{f}} +
  \frac{ \left| V_{{\bf k}} \right|^{2} }{ W_{{\bf k}} }
  \frac{ \partial \langle n_{i}^{f} \rangle }
{\partial \varepsilon_{f}} .
\end{eqnarray}
By inserting Eqs. \eqref{B4} and \eqref{B5} into Eq.~\eqref{B2} one finds
\begin{eqnarray}
  \label{B6}
  0 &=&
  \frac{\partial \tilde{\varepsilon}_{f} }{\partial \varepsilon_{f}}
  \left[
    - \langle n_{i}^{f} \rangle +
    \frac{1}{2}\frac{\nu_{f}}{N}
    \sum_{k\leq k_{F}}
    \left(
      1 + \frac{ \varepsilon_{{\bf k}} - \tilde{\varepsilon}_{f} }{W_{{\bf k}}}
    \right)
  \right] \\
  &&
  + \frac{\partial  \langle n_{i}^{f} \rangle }
{\partial \varepsilon_{f}}
  \left[
    \varepsilon_{f} - \tilde{\varepsilon}_{f} +
    \frac{\nu_{f}}{N}
    \sum_{k\leq k_{F}}
    \frac{ \left| V_{{\bf k}} \right|^{2} }{ W_{{\bf k}} }
  \right]. \nonumber
\end{eqnarray}
We can conclude
\begin{eqnarray}
  \label{B7}
  \langle n_{i}^{f} \rangle &=&
  \frac{1}{2}\frac{\nu_{f}}{N}
  \sum_{k\leq k_{F}}
  \left(
    1 + \frac{ \varepsilon_{{\bf k}} - \tilde{\varepsilon}_{f} }{W_{{\bf k}}}
  \right)
\end{eqnarray}
and
\begin{eqnarray}
  \label{B8}
  \tilde{\varepsilon}_{f} - \varepsilon_{f} &=&
  \frac{\nu_{f}}{N}
  \sum_{k\leq k_{F}}
  \frac{ \left| V_{{\bf k}} \right|^{2} }{ W_{{\bf k}} } .
\end{eqnarray}
Thus the $f$ state occupation of the slave boson mean-field theory 
is reobtained \cite{Fulde}. 
$\left\langle n_{{\bf k}m}^{c}\right\rangle$ can be found in the same way 
\begin{eqnarray}
\label{B9}
  \left\langle n_{{\bf k}m}^{c}\right\rangle &=&
  \frac{1}{2}
  \left(
    1 - \frac{ \varepsilon_{{\bf k}} - \tilde{\varepsilon}_{f} }{W_{{\bf k}}}
  \right)
  \Theta\left( k_{F} - k \right)\\
  &&
  + \frac{\partial \tilde{\varepsilon}_{f}}{\partial \varepsilon_{{\bf k}}}
  \left[
    - \frac{N}{\nu_{f}} \langle n_{i}^{f} \rangle +
    \frac{1}{2}
    \sum_{k^{\prime}\leq k_{F}}
    \left(
      1 +
      \frac{
        \varepsilon_{{\bf k}^{\prime}} - \tilde{\varepsilon}_{f}
      }{
        W_{{\bf k}^{\prime}}
      }
    \right)
  \right] \nonumber \\
  && +
  \frac{ \partial \langle n_{i}^{f} \rangle }
{\partial \varepsilon_{{\bf k}}}
  \left[
    \frac{N}{\nu_{f}}
    \left(
      \varepsilon_{f} - \tilde{\varepsilon}_{f}
    \right) +
    \sum_{k^{\prime}\leq k_{F}}
    \frac{ \left| V_{{\bf k}^{\prime}} \right|^{2} }{ W_{{\bf k}^{\prime}} }
  \right] \nonumber
\end{eqnarray}
By using Eqs. \eqref{B7} and \eqref{B8} one finally obtains
\begin{eqnarray}
  \label{B10}
  \left\langle n_{{\bf k}m}^{c}\right\rangle &=&
    \frac{1}{2}
  \left(
    1 - \frac{ \varepsilon_{{\bf k}} - \tilde{\varepsilon}_{f} }{W_{{\bf k}}}
  \right)
  \Theta\left( k_{F} - k \right) .
\end{eqnarray}

\end{appendix}


\end{multicols}
\widetext

\end{document}